# Brief Report

neuroscience

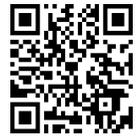

# Nicotinic α7 acetylcholine receptor-mediated currents are not modulated by the tryptophan metabolite kynurenic acid in adult hippocampal interneurons

Peter Dobelis[1*], Andrew L. Varnell[1], Kevin J. Staley[2], & Donald C. Cooper[1]

**The tryptophan metabolite, kynurenic acid (KYNA), is classically known to be an antagonist of ionotropic glutamate receptors. Within the last decade several reports have been published suggesting that KYNA also blocks nicotinic acetylcholine receptors (nAChRs) containing the α7 subunit (α7\*). Most of these reports involve either indirect measurements of KYNA effects on α7 nAChR function, or are reports of KYNA effects in complicated in vivo systems. However, a recent report investigating KYNA interactions with α7 nAChRs failed to detect an interaction using direct measurements of α7 nAChRs function. Further, it showed that a KYNA blockade of α7 nAChR stimulated GABA release (an indirect measure of α7 nAChR function) was not due to KYNA blockade of the α7 nAChRs. The current study measured the direct effects of KYNA on α7-containing nAChRs expressed on interneurons in the hilar and CA1 stratum radiatum regions of the mouse hippocampus and on interneurons in the CA1 region of the rat hippocampus. Here we show that KYNA does not block α7\* nAChRs using direct patch-clamp recording of α7 currents in adult brain slices.**

Kynurenic acid (KYNA) is produced by the metabolism of tryptophan via the kynurenine pathway[1,3]. Classically, KYNA is known for its antagonist actions at ionotropic glutamate receptors, showing the greatest affinity for NMDA-mediated glutamatergic responses[2]. Altered levels of KYNA have been associated with several disease states; increased KYNA levels are seen with Alzheimer's disease, Down's syndrome, and schizophrenia while decreased KYNA levels are associated with end stage Parkinson's and Huntington's disease[3]. Additionally, animal studies indicate that increased brain levels of KYNA are neuroprotective and anti convulsant, while decreased KYNA levels are associated with an increased vulnerability to excitotoxic damage[4].

Another action attributed to KYNA is the antagonism of α7 subunit-containing (α7\*) nicotinic acetylcholine receptors (nAChRs). Several reports from Albuquerque and colleagues present data demonstrating that KYNA also blocks the activation of α7\* nAChRs[4]. However, a recent report from Mok and colleagues examining the effects of KYNA on several different ligand gated ion channels revealed that KYNA had no effect on α7\* nAChRs[6]. Here we present the results of our investigation of KYNA effects on α7 nAChRs expressed on interneurons in the hilar and CA1 stratum radiatum (SR) regions of the mouse hippocampus, as well as α7\* nAChRs expressed on interneurons the rat CA1 SR.

## RESULTS

### Kynurenic acid effects on α7\* nAChRs expressed on mouse hilar interneurons

The results obtained form choline-induced α7\* currents in hilar interneurons are shown in Fig 1. We initially examined the prevalence of α7\* currents in hilar interneurons. Out of the 23 neurons studied, 20 displayed choline-induced and methyllycaconitine (MLA) sensitive whole cell currents characteristic of α7\* nAChRs. Furthermore, in experiments using α7\* null mutant mice, no choline-induced currents were detected (Fig **1a** and **1b**). Next, we examined the effects of KYNA on the choline-induced α7\* currents. In these experiments, stable baseline

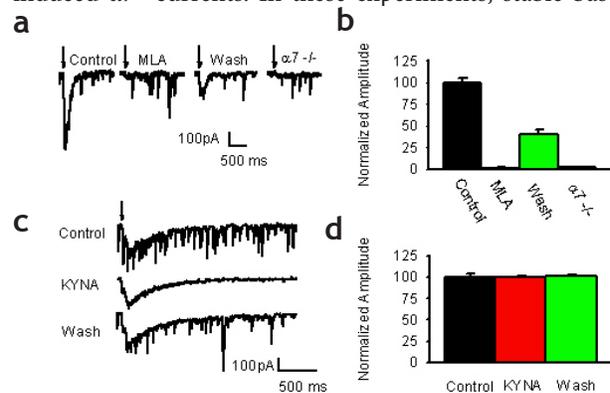

Fig 1: **a.** *Representative traces for the characterization of α7\* currents in mouse hilar interneurons.* The left most trace shows the control response to pressure applied choline. The next two traces show the response to choline in the presence of MLA (10 nM) and after 30 min washout, respectively. The last trace shows the lack of response to choline in α7 null mutant mice. **b.** Methyllycaconitine (MLA) completely blocked the choline response (p = 0.0003 control vs MLA, n = 8), the MLA effect was reversed partially after a washout (p = 0.035 washout vs. MLA, p = 0.005 washout vs control, n = 8), and the choline response for the α7\* null mutant mice differed significantly from wild type (p = 0.0003, n = 8). **c.** *Representative traces for the effect of the bath applied KYNA on choline-evoked α7\* currents.* The top trace shows the control response to pressure applied choline; note the overriding glutamatergic spontaneous EPSCs. The middle trace shows the choline response after a 30 min exposure to 1 mM KYNA; note the absence of the spontaneous EPSCs. The bottom trace shows the response to choline after a 20 min. washout of KYNA; note the reappearance of the spontaneous EPSCs. **d.** KYNA failed to produce any reduction in the choline response (p = 0.97, n = 20).

α7* currents were obtained, followed by bath application of KYNA (1 mM) for 30 min. Choline-evoked α7* currents were then measured after a 20 min washout of KYNA (Fig **1c**). The top trace shows a choline-evoked α7* response with overriding spontaneous glutamatergic-mediated EPSCs. The middle trace shows the choline-evoked α7* response after a 30 min bath application of 1 mM KYNA. The absence of the EPSCs indicates the presence of KYNA. The bottom trace shows the choline-evoked response after a 20 min washout of KYNA. Spontaneous EPSCs returned, indicating the removal of KYNA from the slice (Fig **1d**). Out of the 20 hilar interneurons, KYNA failed to have any significant effect on the choline-evoked α7*-mediated currents.

### *KYNA effects on α7* nAChRs expressed on mouse and rat CA1 SR interneurons*

To determine whether there are regional and/or species differences in α7* nAChR sensitivity to KYNA, we tested the ability of KYNA to block α7* nAChR currents in interneurons in the CA1 SR region in both mice and rats. The representative traces for choline-evoked α7* currents in the mouse CA1 SR appear simular to the KYNA treated cells, indicating no change (Fig **2a**). The top trace shows the control response to pressure applied choline (10 mM) and the bottom trace shows the choline response in the presence of 1 mM KYNA for 30 min. KYNA failed to significantly reduce choline-evoked α7* currents in the five CA1 SR interneurons we tested (Fig **2b**). Representative traces from α7* currents after the pressure applied ACh (1 mM) and KYNA exposure in rat CA1 SR interneurons also showed no change (Fig **2c**). The top trace shows the control response, the middle trace shows the response in the presence of 1 mM KYNA, and the bottom trace shows the response in the presence of the α7-selective antagonist MLA (10 mM). Exposure to a 20 min KYNA bath failed to inhibit α7* currents, while subsequent exposure to MLA produced a greater than 80% blockade of α7* currents (Fig **2d**).

## DISCUSSION

These results clearly indicate that KYNA has no effect on α7* nAChRs in the hilar and CA1 regions of adult mouse and rat hippocampal interneurons. These results are consistent with those reported by Mok and colleagues[6]. Others have concluded that KYNA does interact with the α7* nAChRs using indirect measures of α7* function[4]. KYNA has recently been shown to activate the orphan G-protein receptor GPR-35[1]. GPR-35 is coupled to the $G_{i-o}$ pathway and is expressed throughout the rodent brain[1]. This, in addition to the interactions that KYNA has with other ligand-gated receptors[4] suggests that prior direct α7* mediated physiological effects attributed to KYNA must be made with caution. Further studies are needed to resolve the apparent effects of KYNA on α7* nAChRs. Until then the most direct evidence indicates no role for KYNA on α7* nAChR function.

## METHODS

### *Slice preparation and recordings*

Hippocampal slices were prepared as described by Proctor and colleagues[7]. Whole-cell recordings were made using a standard potassium gluconate based internal solution and a standard ACSF solution. KYNA and MLA were both bath applied to the tissue slices. Both choline and acetylcholine were pressure applied to the slices using established protocols. Expanded methods and recordings can be found at
http://www.neuro-cloud.net/nature-precedings/dobelis/

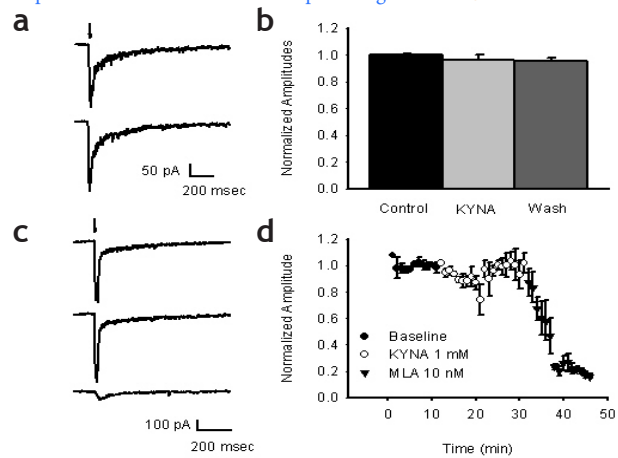

Fig 2. **a.** *Representative traces for choline-evoked α7* currents in mouse stratum radiatum (SR) interneurons*. The top trace is the control response to pressure applied choline (10 mM, 50 ms). The bottom trace shows the choline response after a 30 min. bath exposure to 1 mM KYNA. **b.** KYNA failed to reduce the choline response (p = 0.44, n = 5). **c.** *Representative traces for pressure applied ACh-evoked α7* currents in rat CA1 SR interneurons*. The top trace shows the control response to 1 mM ACh. The middle trace shows the ACh respose after the 30 min. bath exposure to 1 mM KYNA. the bottom trace shows the ACh response after a 20 min. bath exposure to 10 nM MLA. **d.** The last five traces were averaged for each condition: control, KYNA, and MLA. KYNA failed to block the ACh-induced α7* currents (p = 0.71) while MLA significantly blocked the response (p = 0.0001, n = 5).

## ACKNOWLEDGEMENTS

This work was funded and supported by the American Epilepsy Foundation/Milken Family Foundation Fellowship (P.D.), R21 DA026918 (J. Stitzel), National Institute on Drug Abuse grant R01-DA24040 (to D.C.C.), NIDA K award K-01DA017750 (to D.C.C.).

## PROGRESS AND COLLABORATIONS

To see up to date progress on this project or if you are interested in contributing to this project visit: http://www.neuro-cloud.net/nature-precedings/dobelis/

## AUTHOR CONTRIBUTIONS

P.D., D.C.C. & K.J.S. designed the experiments. P.D. recorded and analyzed the data. P.D., A.L.V. & D.C.C. wrote and prepared the manuscript.

Submitted online at http://www.precedings.nature.com